\documentclass{article}
\usepackage{amsmath}
\usepackage{caption}
\usepackage{indentfirst}
\usepackage{booktabs}
\usepackage{multirow}
\usepackage{float}
\newtheorem{rmk}{Remark}

\usepackage{titlesec}
\titlespacing*{\section}{0pt}{\dimexpr\baselineskip-50pt}{\dimexpr\baselineskip-50pt}
\usepackage{PRIMEarxiv}
\usepackage[utf8]{inputenc} % allow utf-8 input
\usepackage[T1]{fontenc}    % use 8-bit T1 fonts
\usepackage{hyperref}       % hyperlinks
\usepackage{url}            % simple URL typesetting
\usepackage{booktabs}       % professional-quality tables
\usepackage{amsfonts}       % blackboard math symbols
\usepackage{nicefrac}       % compact symbols for 1/2, etc.
\usepackage{microtype}      % micro typography
\usepackage{fancyhdr}       % header
\usepackage{graphicx}       % graphics
\graphicspath{{media/}}     % organize your images and other figures under the media/ folder

\title{Introducing Hybrid Modeling with Time-series-Transformers: A Comparative Study of Series and Parallel Approach in Batch Crystallization
}

\author{
  Niranjan Sitapure \\
  Dept. of Chemical Engineering \\
  Texas A\&M University \\
  College Station, TX 77801\\
  \texttt{niranjan\_sitapure@tamu.edu} \\
   \And
  Joseph Sang-Il Kwon$^{*}$\\
  Dept. of Chemical Engineering \\
  Texas A\&M University \\
  College Station, TX 77801\\
  \texttt{kwonx075@tamu.edu} \\
}
\begin{document}

\captionsetup[figure]{font=small,skip=-5pt}
\captionsetup[table]{font=small,skip=5pt}
\maketitle

\begin{abstract}
	
Most existing digital twins rely on data-driven black-box models, predominantly using deep neural recurrent, and convolutional neural networks (DNNs, RNNs, and CNNs) to capture the dynamics of chemical systems. However, these models have not seen the light of day, given the hesitance of directly deploying a black-box tool in practice due to safety and operational issues. To tackle this conundrum, hybrid models combining first-principles physics-based dynamics with machine learning (ML) models have increased in popularity as they are considered a `best of both worlds’ approach. That said, existing simple DNN models are not adept at long-term time-series predictions and utilizing contextual information on the trajectory of the process dynamics. Recently, attention-based time-series transformers (TSTs) that leverage multi-headed attention mechanism and positional encoding to capture long-term and short-term changes in process states have shown high predictive performance. Thus, a first-of-a-kind, TST-based hybrid framework has been developed for batch crystallization, demonstrating improved accuracy and interpretability compared to traditional black-box models. Specifically, two different configurations (i.e., series and parallel) of TST-based hybrid models are constructed and compared, which show a normalized-mean-square-error (NMSE) in the range of $[10, 50]\times10^{-4}$ and an $R^2$ value over 0.99. Given the growing adoption of digital twins, next-generation attention-based hybrid models are expected to play a crucial role in shaping the future of chemical manufacturing. 
\end{abstract}
% keywords can be removed

\section{Introduction}

% Increase the spacing between the caption and text
\setlength{\belowcaptionskip}{10pt}
\setlength{\parindent}{10pt} % Default is 15pt.

The development and implementation of digital twins in the chemical industry have experienced significant growth in recent years, driven by the transition to Industry 4.0 coupled with the advancements in machine learning (ML) and enterprise-based artificial intelligence (AI) \cite{venkatasubramanian2019promise, pistikopoulos2021process}. These digital twins serve as virtual replicas of physical systems, allowing for enhanced online process monitoring, control, and optimization. By integrating real-time data, advanced analytics, and ML/AI algorithms, digital twins enable continuous monitoring, predictive analysis, and optimization of chemical processes in real-time \cite{hwang2022model, schweidtmann2021machine, bhadriraju2023adaptive}. A majority of the existing digital twins utilize data-driven black-box models, which are majorly based on different ML architectures as they provide high computational efficiency. Specifically, deep, recurrent, or convolution neural networks (DNN, RNN, and CNN) are used to mimic the process dynamics of complex chemical systems \cite{schweidtmann2021machine, venkatasubramanian2022artificial, pan2022data, sitapure2023require}. For instance, various RNN and DNN models have been utilized in the literature to mimic the crystallization of pharmaceutical and food products by Wu and colleagues \cite{wu2019machine, wu2019machineII, zheng2022machine, zheng2022online}. Similarly, Kwon and colleagues have DNN-based models for continuous crystallization of quantum dot (QD) systems. \cite{SITAPURE2020127905, sitapure2021cfd}. Similarly, other DNN-based models have also been demonstrated for modeling and control of thin-film deposition for different substrates \cite{kimaev2019nonlinear, kimaev2020artificial, sitapure2022neural}. Furthermore, Braatz and colleagues have demonstrated a plethora of impressive different ML models for  the prediction of battery life, and developing optimization or control frameworks using these ML models \cite{severson2019data, attia2020closed, schaeffer2023machine}. 

Unfortunately, despite the accuracy of the above ML models, the deployment of such black-box models in practical industrial settings has been limited. One major concern is the hesitance surrounding the direct implementation of these models due to safety and operational issues. Deploying a black-box tool that lacks interpretability may pose risks in critical applications where human safety and system reliability are paramount. The second concern is that these models do not integrate any prior knowledge of various chemical systems (e.g., mass and energy balance equations (MEBEs), population balance model (PBM), known Arrhenius-type kinetics, etc.), thereby leaving out the known system dynamics. Recently, physics-informed neural networks (PINNs) have provided a certain level of integration between \textit{a priori} system knowledge and various machine learning techniques. Notably, Karniadakis and colleagues have developed various PINNs for describing complex transport problems, such as heat transfer, Navier-Stokes, diffusion systems, among others, and multiple process systems described by different forms of partial differential equations (PDEs) \cite{raissi2019physics, lin2021operator, karniadakis2021physics, goswami2022physics, yu2022gradient, jagtap2022physics}. That said, although PINNs consider continuum equations in the residual function, they do not explicitly integrate the dynamics of complex nonlinear chemical systems (e.g., PBM in crystallization, multi-reaction kinetics in fermentation, or catalysis systems) with data-driven components. An alternative approach to PINNs is the method of sparse identification of nonlinear system dynamics (SINDy) and operable adaptive sparse identification of systems (OASIS). these methods employ a library of basis functions to identify a sparse set of equations that describe dynamic chemical processes \cite{Bhadriraju2019,de2020pysindy, bhadriraju2021oasis}. Essentially, SINDy and OASIS allow utilization of \textit{a priori} physics, which underlie the system dynamics, into their library of basis functions. For instance, given the well-known fact that reaction kinetics follow an Arrhenius-type exponential temperature dependence and power-law dependence on concentration, various combinations of these variables can be used as basis functions \cite{Bhadriraju2019, bhadriraju2020operable}. These techniques have been successfully implemented in various complex systems, including battery systems, fault prognosis, and even resilience assessment in reactor systems \cite{bhadriraju2021risk, pawar2023resilience}. Additionally, Brunton and colleagues have conducted comprehensive theoretical and applied research studies, demonstrating the capabilities of SINDy models in control, optimization, and modeling for a variety of applications \cite{brunton2016sparse, mangan2016inferring, kaiser2018sparse}. It is worth noting, however, that while SINDy and OASIS incorporate some \textit{a priori} knowledge with process data, their training approach essentially solves a regression problem to find the coefficient of basis function that provides a combined representation of the process dynamics. As a result, these techniques operate more like grey-box model. 

Thus, in recent times various types of hybrid models that provide more direct integration between system-agnostic first principles with system-specific data-driven parameters have been considered as a `best-of-both-worlds' approach. \cite{chaffart2018optimization, ghosh2019hybrid, lee2020development, lee2020identification, hassanpour2020hybrid, patel2020integrating}. Basically, a first-principle (FP) module comprises typical MEBEs, rate kinetics, and PBM (if applicable) while the data-driven part estimates the system-specific parameters (i.e., kinetic rate constant, selectivity, separation coefficient, etc.), thereby acting as an input to the FP module. For instance, Kwon and his colleagues have developed a plethora of different hybrid models for the modeling and control of fermentation, hydraulic fracturing, and biological processes \cite{bangi2020deep, bangi2022physics, shah2022deep, shah2023multi, bangi2023deep}. Specifically, the current system states (e.g., concentration, temperature, flow rate, etc.) are utilized to predict the kinetic parameters (i.e., rate constants, leak-off rate, and others), which are then subsequently integrated with the first-principled reactor model or hydraulic fracking model. These models use typical DNNs, which are not adept at predicting complex time-varying parameters that are ubiquitous in complex chemical processes \cite{hamzaccebi2008improving, shin2021short, ozbek2021prediction}. Also, the DNNs are not capable of utilizing contextual information on the trajectory of the process dynamics (e.g., cooling curves in batch crystallization or decaying oxygen concentration in fermentation reactors). Thus, there is a need for an alternative hybrid modeling approach that (a) leverages available process data to approximate the underlying kinetic function, allowing it to capture the complex behavior of the system; and (b) accurately captures the short-term and long-term evolution of system states.  

Recently, a few notable works have demonstrated the use of attention-based time-series-transformers (TSTs) and modified NLP-transformers for modeling and control of different chemical processes\cite{vogel2023learning,sitapure2023exploring}. For example, Kang and colleagues developed the MOFTransformer, which incorporates atom-based graphs and energy-grid embedding to capture both local and global features of metal-organic-frameworks (MOFs) structures \cite{kang2023multi}. Also, Venkatasubramanian and Mann utilized the SMILES representation of different chemical precursors to predict potential reaction products \cite{mann2021predicting}. More pertinent to time-series modeling, Kwon and Sitapure showcased CrystalGPT, which is a unified digital twin for 20+ different sugar crystal systems, and it is approximately 10 times more accurate than other state-of-the-art (SOTA) ML methods \cite{sitapure2023crystalgpt}. Specifically, all the above transformer-based models show remarkable predictive performance by virtue of the multiheaded attention mechanism, positional encoding (PE), and parallelization-friendly architecture. Basically, the attention mechanism performs a scaled-dot product calculation between
different input tensors, enabling it to selectively assign higher attention scores to time-steps that show significant process changes. Further, these transformer models utilize current process data (i.e., temperature, concentration, suspension density, etc.) for the current and past $W$ time-steps (i.e., a window size of $W$), to understand the long-term and short-term changes in process states, thereby gaining a contextual understanding of the overall process dynamics \cite{vaswani2017attention, devlin2018bert}. 

\begin{figure}[!ht]
	\begin{center}
		\centerline{\includegraphics[width=1\columnwidth]{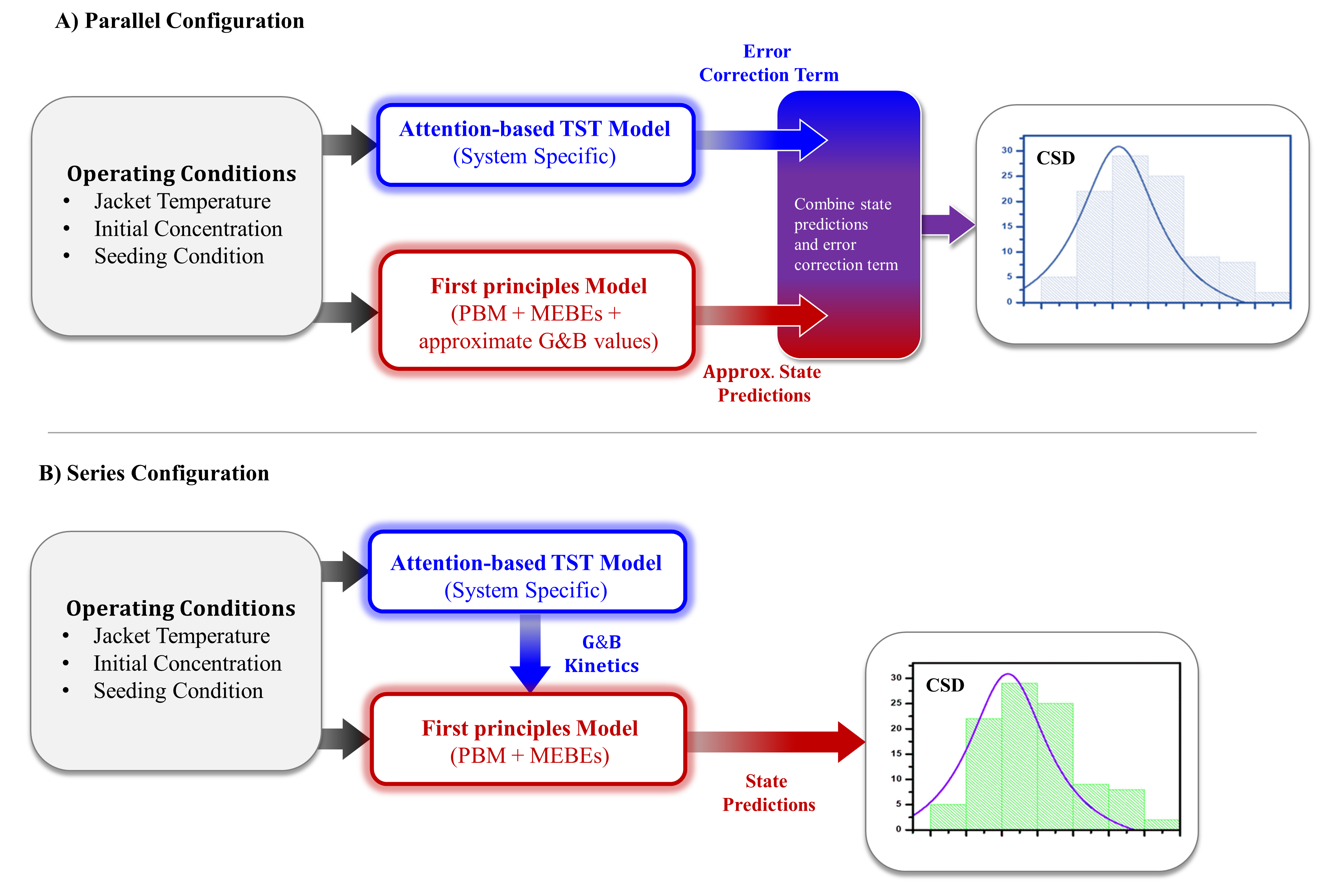}}
	\end{center}
	\caption{Schematic illustration of parallel and series configuration for TST-based hybrid model.}
	\label{configuration_schematic}
\end{figure}

Thus, to tackle the abovementioned two challenges, a first-of-a-kind TST-based hybrid framework was constructed for a complex nonlinear chemical process of batch crystallization. Here, a scenario encountered frequently by the industry was considered: (a) Process data (e.g., concentration, temperature, process yield, crystal moments, etc.) is available, and (b) there is an initial idea about the crystallization kinetics (e.g., growth (G) and nucleation rates (B)) for the desired crystal system or a previously studied similar crystal system. In this work, we utilized the aforementioned information to develop two unique TST-based hybrid modeling approaches, which we refer to as the parallel and series approaches, as illustrated in Figure~\ref{configuration_schematic}. In the parallel approach, an FP model, which includes PBM and MEBEs, is used in conjunction with approximate $G$\&$B$ kinetics to simulate the evolution of system states. Here, the approximate kinetics could either be based on a previously studied crystal system with similar characteristics or be a rough estimate of the $G$\&$B$ kinetics. Regardless, these kinetics would exhibit a certain degree of deviation from the true kinetics of the desired crystal system, resulting in a plant-model mismatch. To rectify this, we introduce a TST model that takes current process data (e.g., temperature, concentration, and other states) as input and produces error correction terms. These terms are then added to the state predictions by the FP model, creating an integrated parallel configuration, as shown in Figure\ref{configuration_schematic}A. For the series approach, the current state evolution is input into a TST model. The TST model learns to estimate the instantaneous $G$\&$B$ kinetics accurately, which are then integrated with the FP model (comprising of PBM + MEBEs). The result is a direct prediction of state evolution in batch crystallization, depicted in Figure~\ref{configuration_schematic}B. Although both the approaches take similar inputs, and yield the same outputs, the training and testing methodologies are starkly different. Most notably, although there is an amalgamation of data-driven aspects and an FP model in the parallel configuration, the error correction terms are predicted via a black-box TST model. On the contrary, the fully-integrated, series approach is considered a `true-hybrid' approach as it utilizes the process data to learn to estimate true crystallization kinetics (i.e., $G$\&$B$), thereby achieving more interpretability. Next, to compare these two approaches, industrial batch crystallization of sugars is considered over 500 different operating conditions, and the simulation results showcase that the series approach shows 5 times less deviation from the true system dynamics than the parallel approach. Further, the state predictions by the series approach have an average $R^2$ over 0.99 while boosting a combined normalized-mean-squared-error (NMSE) in the range of $[10, 50]\times10^{-4}$. Overall, given the growing inclination for adopting digital twins in the chemical industry, such next-generation attention-based hybrid models are expected to play a crucial role in shaping the future of chemical manufacturing.

The rest of this manuscript is organized as follows: we begin with a concise mathematical representation of batch crystallization, and explain the representative case study. This is followed by an exploration of the internal computations and training protocols for TST and TST-based hybrid models. Then, model validation and a comparison between the two approaches are presented. The paper concludes with a discussion and summary of the findings from our research.  

%\section{Constructing Hybrid Models}

\section{First Principles Modeling of Batch Crystallization}

As mentioned earlier, industrial batch crystallization of sugar systems is considered as a representative case study of a complex non-trivial chemical system. Here, it is assumed that there is availability of process data (i.e., concentration, temperature, crystal moments) from past operation conditions, and a system-agnostic FP module consisting of PBM and MEBEs is known. For instance, the crystal growth rate and nucleation rate serve as primary descriptors of the kinetics of a crystal system. A generalized growth rate equation for most of the crystal systems is given as follows: 

\begin{equation}
	\begin{split}
		G~(m/s) = k_{\scriptscriptstyle g}(S-1)^{\scriptscriptstyle P_{\scriptscriptstyle g}}
	\end{split}\label{G_rates}
\end{equation}
where $R$ is the universal gas constant, $T$ is the temperature, and $S$ is the supersaturation. The kinetic parameters are denoted as $k_g$ and   $p_g$. Similarly, a typical nucleation rate ($B$) is given in the following manner:
\begin{equation}
	\begin{split}
		B~(\#/kg \cdot s) = k_b(S-1)^{\scriptscriptstyle P_{\scriptscriptstyle b}}\label{B_rates}
	\end{split}
\end{equation}
where $k_b$ and $p_b$ represent kinetic parameters. Now there can be two cases for modeling batch crystallization of a certain new system. In the first case, approximate kinetics (i.e., $[k_g,k_b,p_g,p_b]$ values) for a very similar system might be known that can be considered as approximate kinetics. For instance, many food product manufacturers will operate various sugar systems that are somewhat similar to one another \cite{sitapure2023crystalgpt} (e.g., aspartame and a modified more potent version of aspartame). Thus, the approximate kinetics can be utilized as a proxy for the actual kinetics, and adequate predictions of state evolution can be generated. In the second case, the information about $[k_g,k_b,p_g,p_b]$ values is not known but a general idea about the magnitude of these parameters can be estimated. Generally speaking, the supersaturation dependence of most crystal systems (i.e., $p_g$ and $p_b$) is between 0 and 3. Similarly, by performing a basic analysis of the existing process measurements (e.g., average crystal size), a rough estimate of the growth rate can be considered. For example, the crystallization of dextrose is a very slow process that is performed over 24 hr, which leads to a crystal growth of around 100 to 200 microns, thereby suggesting $G \in [5,10]$ $\mu m$/hr). In this scenario, a hybrid model that can help estimate the actual $G$\&$B$ kinetics is required. 

Once $G$\&$B$ kinetics are known, the size distribution of crystals and their temporal evolution can be traced using a PBM that utilizes a population density function, $n(L,t)$, and is given as follows \cite{worlitschek2004model}:

\begin{equation}
	\begin{aligned}
		\frac{\partial n (L,t)}{\partial t} + \frac{\partial (G(T,C_s)n(L,t))}{\partial L} = B(T,C_s)
	\end{aligned}\label{PBM}
\end{equation}
where $n (L,t)$ represents the number of crystals of size $L$ at time $t$, $B(T,C_s)$ is the total nucleation rate, and $G(T,S)$ represents the crystal growth rate. Then, the PBM is integrated with MEBEs, which are presented below:
\begin{equation}
	\begin{gathered}
		\frac{d C_s}{d t} = -3\rho_ck_vG\mu_2 \\
		mC_p\frac{dT}{dt} = -UA(T - T_{j}) - \Delta H\rho_{c}3k_vG\mu_2
	\end{gathered}\label{MEBEs}
\end{equation}
where $\mu_2$ is the second moment of crystallization, $k_v$ is the shape factor, $\rho_c$ is the crystal density, $C_p$ is the heat capacity of the crystallization slurry, $m$ is the total mass of the slurry, $UA$ is the area-weighted heat transfer coefficient, and $\Delta H$ is the heat of crystallization. For a more comprehensive description of modeling crystallization systems, the reader is referred to relevant literature studies \cite{braatz2002advanced,nagy2012advances,kwon2013modeling,kwon2014crystal,sitapure2023unified, ochsenbein2014crystallization}.

\section{Time-series-Transformers (TSTs)}
\subsection{Working of Encoder-Decoder Transformers}
A sophisticated vanilla transformer architecture employed for natural language processing (NLP) tasks is characterized by encoder/decoder blocks, housing identical sub-layers, and featuring a globally pooled output layer. The input sequence undergoes a series of transformations encompassing truncation/padding, tokenization, positional encoding, self-attention, and cross-attention mechanisms, ultimately leading to the contextualized embedding of the source sequence. Importantly, the initial input tensor ($X_{RAW}$) is lifted into a higher-dimensional space ($d_{model}$) and subsequently processed through a sinusoidal positional encoder (PE). This encoder systematically identifies each word's position within the sequence, thereby producing the tensor $X_{PE}$. Following this, $X_{PE}$ is dispatched to a stack of encoder blocks, wherein each block employs multiple attention heads (MAHs) to diligently compute `self-attention' scores, subsequently subjected to processing by a feed-forward network (FFN). The computation of attention scores by each individual attention head is instrumental in capturing the semantic relationships, context, and relative importance of every word, thereby culminating in the formation of contextually enriched embeddings. It is common practice to utilize the Query-Key-Value ([\textbf{Q},\textbf{K},\textbf{V}]) approach for calculating attention scores \cite{wen2022transformers}, with the results amalgamated from the diverse MAHs to generate the tensor $X_{EN}$. Next, $X_{EN}$ is forwarded to a stack of decoder blocks, with each block having its own set of MAHs and independent input ($X_{DEC}$) to facilitate the computation of `cross-attention'. This comprehensive framework enhances the transformer's ability to focus on high-value cross-attention scores. This enhancement facilitates a seamless transition and connection between the source and target sequences of words, ultimately resulting in the generation of text that closely mirrors human-like communication. For a more detailed understanding, we encourage readers to explore the existing literature on transformer models\cite{vaswani2017attention, devlin2018bert, brown2020language}.

\subsection{TST Architecture}
\begin{figure}[!ht]
	\begin{center}
		\centerline{\includegraphics[width=0.65\columnwidth]{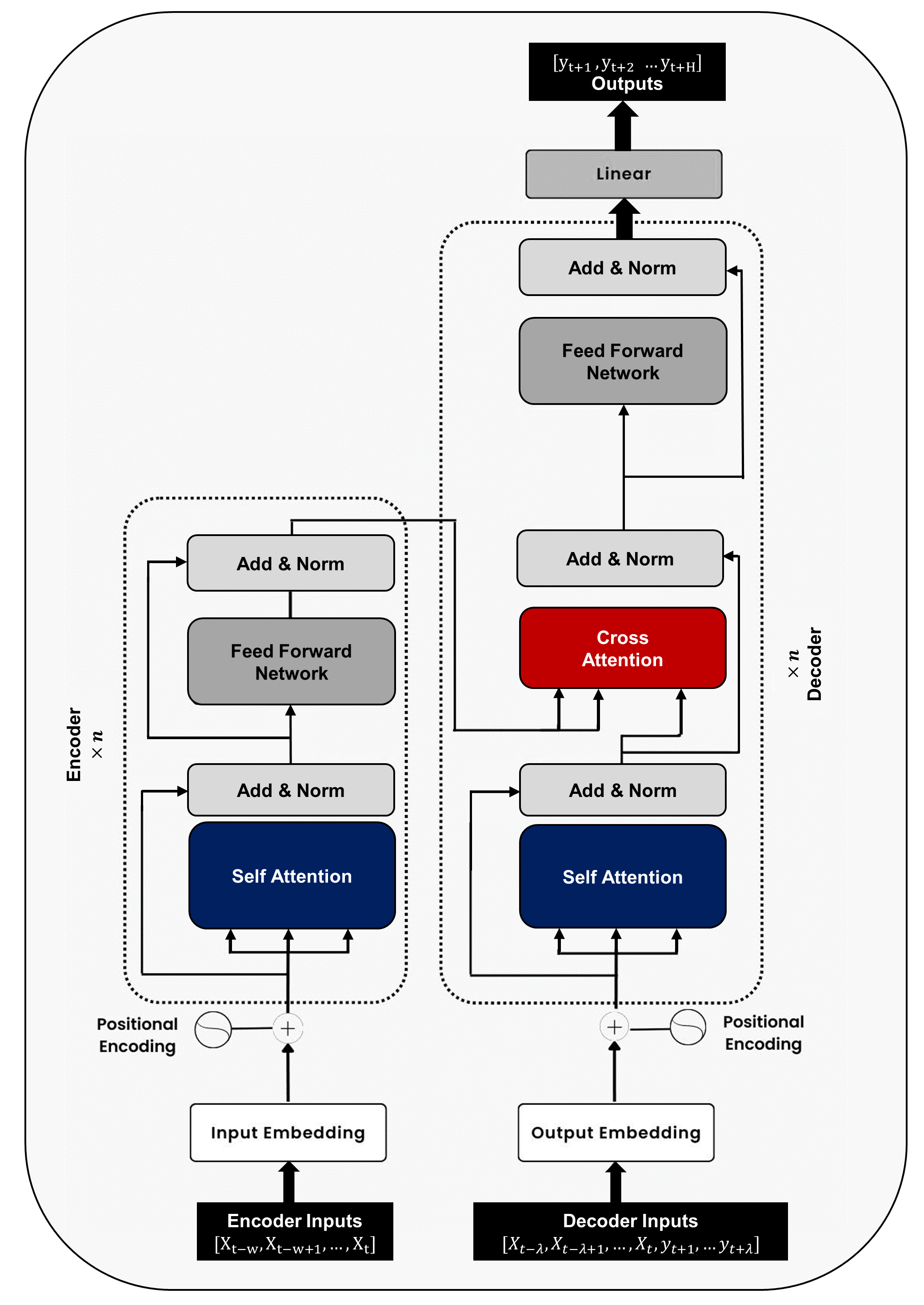}}
	\end{center}
	\caption{Generalized architecture for an encoder-decoder TST.}
	\label{TST_schematic}
\end{figure}

Taking inspiration from the above-mentioned encoder-decoder architecture of NLP-transformers, a novel TST framework that is specifically tailored for time-series prediction tasks was developed in our previous works \cite{sitapure2023exploring, sitapure2023crystalgpt}. A brief explanation of this architecture is presented here. First, time-series data from chemical systems, which consists of state information spanning the current and the past $W$ time-steps (i.e, the source sequence), is preserved in a k-dimensional input tensor, represented as $[X_{t-W}, X_{t-W+1} ... X_{t}]$. Next, it is lifted to a higher-dimensional space ($d_{model}$), representing the internal hidden dimension of a TST. Second, PE terms based on the time variable $t_i$ can be tracked, and used as follows:

\begin{equation}
	\begin{aligned}
		& Even~Position: PE_{(t_i,2j)} = sin\left(\frac{t_i}{10000^{2j/d}}\right) &\\
		& Odd ~Position: PE_{(t_i,2j+1)} = cos\left(\frac{t_i}{10000^{2j/d}}\right) & 
	\end{aligned}\label{modified_positional_encoding}
\end{equation}
where $t_i$ is the time at location $i$ in the input sequence, and $j \in \mathbb{R}^{d_{model}}$ is the feature dimension. By incorporating PE, the transformer adeptly generates sine and cosine waves with varying wavelengths. This capability allows it to attend to different positions in the input sequence with a nuanced understanding, thereby enhancing its performance \cite{sitapure2023exploring, vaswani2017attention}. Each distinct input value at position $i$ and dimension $j$ is given a unique identifier, either as $PE_{(i,2j)}$ or $PE_{(i,2j+1)}$, depending on whether the time-step occupies an even or odd position within the source sequence. Furthermore, we can perceive the output of the TST as a sequence with a prediction horizon of $H$ and dimension $v$, represented as $[y_{t+1}, y_{t+2}, ... y_{t+H}]$. This sophisticated framework, enriched with various parameters, empowers the transformer to discern and adapt to temporal dependencies, ultimately providing a powerful sequence-to-sequence prediction capability. The TST architecture, as described schematically in Figure~\ref{TST_schematic}, is comprised of several encoder-decoder blocks, each housing \textit{{n}} attention-heads. 

Furthermore, the attention mechanism in each of the attention heads is based on the [\textbf{Q},\textbf{K},\textbf{V}] mechanism, which symbolizes the system's state at position $i$ (i.e., $X_{t-i} \in [X_{t-W}, X_{t-W+1} ... X_{t}]$) with an observation horizon $W$ (i.e., similar to a sentence). Consequently, the internal computations within an attention layer are presented as follows:   

\begin{equation}
	\begin{aligned}
		& A_{P,n} = \sum_{i}^{k} \lambda_{n,i} \textbf{V}&\\
		& \lambda_{n,i} = \frac{exp\left(\textbf{Q}^T\textbf{K}_i/\sqrt{D_k}\right)}{\sum_{j=1}^{k}exp\left(\textbf{Q}^T\textbf{K}_j/\sqrt{D_k}\right)} & \\ 
		& \sum_{i=1}^{k} \lambda_{n,i} =1& 
	\end{aligned}\label{QKV_model}
\end{equation}
where $A_{P,n}$ represents the attention value for head $n$ in encoder block $P$. The elements \textbf{Q}$\in \mathbb{R}^{D_k}$, \textbf{K}$\in \mathbb{R}^{D_k}$, and \textbf{V}$\in \mathbb{R}^{D_v}$ denote queries, keys, and values, respectively, where $D_k$ and $D_v$ are the dimensions of the keys and values, respectively. Additionally, the attention score ($\lambda_{n,i}$) indicates the relative importance between different words in the input sequence. The softmax calculation in Eq.~(\ref{QKV_model}) ensures that the scaled sum of attention scores equals 1 for each input (i.e., $\sum_{i=1}^{k} \lambda_{n,i}$ =1). It is important to understand that during the self-attention process within the encoder block, \textbf{Q, K} and \textbf{V} all derive from the same input tensor, $X_{PE}$. In contrast, during the cross-attention calculation in the decoder block, \textbf{Q} stems from $X_{DEC}$, whereas \textbf{K} and \textbf{V} originate from the processed output of the encoder stack, denoted by $X_{EN}$. Furthermore, MAHs are trained on the same [\textbf{Q},\textbf{K},\textbf{V}] to facilitate automatic learning of diverse features from the input data. The output from MAHs, therefore, represents a combination of these different features. This output is represented as follows: 

\begin{equation}
	MAH(\textbf{Q},\textbf{K},\textbf{V}) = \text{Concat}[head_1, head_2, ...,  head_n]
	\label{multiheadattention}
\end{equation}
Next, the attention scores obtained from each encoder or decoder block undergo a positional encoding process, illustrated as follows:  

\begin{equation}
	FFN(\sigma_{i+1}) = ReLU(\sigma_i \theta_i + b_i)
	\label{feedforward}
\end{equation}
where $\sigma_i $ represents the intermediate state from previous layers, and $\theta_i$ and $b_i$ are trainable parameters of the neural network. For a more comprehensive understanding of the aforementioned computations, readers are referred to the literature  \cite{sitapure2023exploring, sitapure2023crystalgpt, wen2022transformers, zeng2022transformers}.

\section{Construction of Hybrid Models}

As mentioned earlier, two different approaches toward TST-based hybrid models will be evaluated in this work. The requirements, construction, and training methodology for these two approaches are explained in this section.

\subsection{Problem Description}\label{problem_description}

To demonstrate the working and performance of TST-based hybrid models, the batch crystallization of dextrose is considered as a representative case study. The objective of the hybrid models is to have accurate state predictions that minimize the plant-model mismatch. Two types of information are available to do this task. First, there is availability of process data (i.e., $X (t) \in $[$T_j(t), C_s(t), T(t), \bar{L}(t), \mu_0(t), \mu_1(t), \mu_2(t), \mu_3(t), t$]) for over 500 operating conditions, which is generated using the true kinetics (Equation~\ref{dextrose_kinetics}). Second, a very common scenario faced by industrial manufacturers is considered, wherein it assumed that the true kinetics of dextrose crystallization is not known, but an approximate of the crystallization kinetics is available as shown below:

\begin{equation}
	\begin{gathered}
		G_{approx}~(m/s) = 1\times10^{-8} (S-1)^{2} \\ 
		B_{approx}~(\#/kg \cdot s) = 1 \times 10^{5}(S-1)^{2}
	\end{gathered}\label{approx_kinetics}
\end{equation}
Here, the approximate kinetics for a new crystal system can be obtained through two approaches: (a) utilizing kinetics from an existing system with similarities to the new one, or (b) conducting basic experimental studies to develop a rough estimate. For instance, a food manufacturer has newly installed a batch crystallizer for dextrose crystallization and does not know the exact know $G$\&$B$ kinetics. However, kinetic data from other crystallizers for fructose sugars, maltose, or other similar sugars are available. Given the similarity of dextrose with these crystal systems, the kinetics of fructose/maltose/other can be treated as approximate kinetics for providing some baseline predictions. Indeed, while the state predictions from these simulations may not precisely match the true system dynamics, they serve as baseline predictions, providing a general idea about state dynamics and temporal evolution. Alternatively, researchers can conduct low-resource-intensive experimental studies or pilot runs to get a rough estimate of the $G$\&$B$ kinetics of dextrose. Next, utilizing the above approximate kinetics in the FP model (PBM + MEBEs), batch crystallization can be simulated. Since the true kinetics are not known, there will be a plant-model mismatch between the state predictions and the true system dynamics. To tackle this challenge, two different hybrid modeling approaches (i.e., series and parallel) are considered in this work. In the series hybrid model, the objective is to use the TST model to estimate the kinetic parameters $[k_g,k_b,p_g,p_b]$ that will minimize the plant model mismatch and the approximate kinetics can be used as an initial condition as shown in Figure~\ref{configuration_schematic}.  In the parallel configuration, the TST model is utilized to estimate the error correction terms that can be directly added to the baseline predictions (i.e., state predictions generated using the approximate kinetics) that will minimize the plant-model mismatch. 

Furthermore, two architectures of TST models (i.e., \textbf{Base} and \textbf{Large}) that are within each of the hybrid models were considered. More precisely, a TST model has several hyperparameters that need to be chosen such as the number of encoders/decoders, the internal dimensions ($d_{model}$), the size of FFN, and other aspects. Based on our previous experience with a plethora of TST models, \textbf{Base} and \textbf{Large} architectures were considered in this work, and their details are mentioned in Table~\ref{model_arch}. Although more architectural variants (e.g., Small and Mega) of the TST model can be considered, for the sake of demonstration, only a \textbf{Base} and \textbf{Large} model is considered. Basically, these hyperparameters of the TST model affect the width and depth of a TST \cite{Levine2020Limits}. Specifically, the width of a transformer model is majorly influenced by the number of attention heights and the size of the FFN. A larger number of attention heads allow for the capture of prominent and subtle system interdependencies, which can be pooled together for an improved understanding of the input/output relationship. On the contrary, the depth of a TST model is determined by the number of encoder/decoder blocks. Given that every encoder/decoder block consists of $n$ attention heads in parallel, which are connected to a simple FFN layer, increasing the number of encoder/decoders allows each successive FFN to capture more nonlinearity in the input/output relationship  \cite{Xu2019Lipschitz}. Thus, the width and depth of TSTs are critical hyperparameters that significantly influence the TST model's ability to accurately learn to correlate the current state information with the kinetic parameters of $G$\&$B$ (in the case of the series hybrid model), and learn to correlate the error correction terms (in the case of the parallel-hybrid model). More details about the effect of depth and width of TST models can be found in the literature \cite{sitapure2023exploring, sitapure2023crystalgpt, Levine2020Limits}. 

\begin{table}[H]
	\caption{\label{model_arch}
		Architectural details of the two TST models that were considered in this work. }
	\centering
	\begin{tabular}{@{}ccc@{}}
		\toprule
		\textbf{}                   & \textbf{Base} & \textbf{Large} \\ \midrule
		\# Parameters      & $\sim$ 2.5M                  & $\sim$ 6.5M                 \\
		\# Attention heads & 4                    & 8                   \\
		$d_{model}$          & 128                   & 256                 \\
		\# Encoder blocks  & 2                    & 6                   \\
		\# Decoder blocks  & 2                    & 6                   \\
		\# Neurons in FFN  & 256                   & 256                 \\ \bottomrule
	\end{tabular}
\end{table}

\subsection{Acquisition of Process Data} 

To effectively demonstrate the predictive capabilities of the hybrid model configurations, a comprehensive compilation of process data becomes important. In this regard, the well-studied system of dextrose crystallization serves as the ideal candidate. The dataset is constructed by recording the state evolution for batch crystallization of dextrose using known $G$\&$B$ kinetics from the literature \cite{markande2012influence}. In simpler terms, since industrial process data for a batch crystallizer is not available, a comprehensive simulation (i.e., a virtual experiment) of dextrose crystallization using exact growth and nucleation kinetics is utilized as synthetic process data for model training and testing. 
More precisely, an exhaustive set of over 500 operating conditions are considered, encompassing various key variations for each simulation: an arbitrarily selected (a) jacket cooling curve with temperatures ranging from $5^\circ$C to $40^\circ$C; (b) solute concentration spanning from $0.6$ to $0.9$ kg/kg; and (c) seed loading comprising of percentages ranging from $0\%$ to $20\%$ (measured in \%w/w), accompanied by a diverse range of seed crystal sizes varying between $10~\mu m$ to $150~\mu m$. Moreover, the crystallization kinetics of dextrose is available in the literature, which will be referred to as true kinetics for the rest of the manuscript, and is given below \cite{markande2012influence}: 

\begin{equation}
	\begin{gathered}
		G~(m/s) = 1.14 \times 10^{-3} \exp\left(\frac{-29549}{RT}\right) (S-1)^{1.05} \\ 
		B~(\#/kg \cdot s) = 4.50 \times 10^{4}(S-1)^{1.41}
	\end{gathered}\label{dextrose_kinetics}
\end{equation}

The combined dataset was divided into 35K data points for the training set, 15K for the validation set, and another 10K for the testing set. Each data point contains information comprising of 9 process states (i.e., [$T_j, C_s, T, \bar{L}, \mu_0, \mu_1, \mu_2, \mu_3, t$]) for both the current and previous $W$ time-steps. This extensive variation in process conditions facilitates data generation across all possible operating regimes, resulting in a comprehensive dataset that aids the series hybrid model to estimate the kinetic parameters, and for the parallel hybrid model to learn the error correction terms to minimize plant-model mismatch. Basically, the objective of the hybrid model is to make accurate model predictions without fully knowing the true $G$\&$B$ of dextrose crystallization.

\begin{figure}[!ht]
	\begin{center}
		\centerline{\includegraphics[width=0.85\columnwidth]{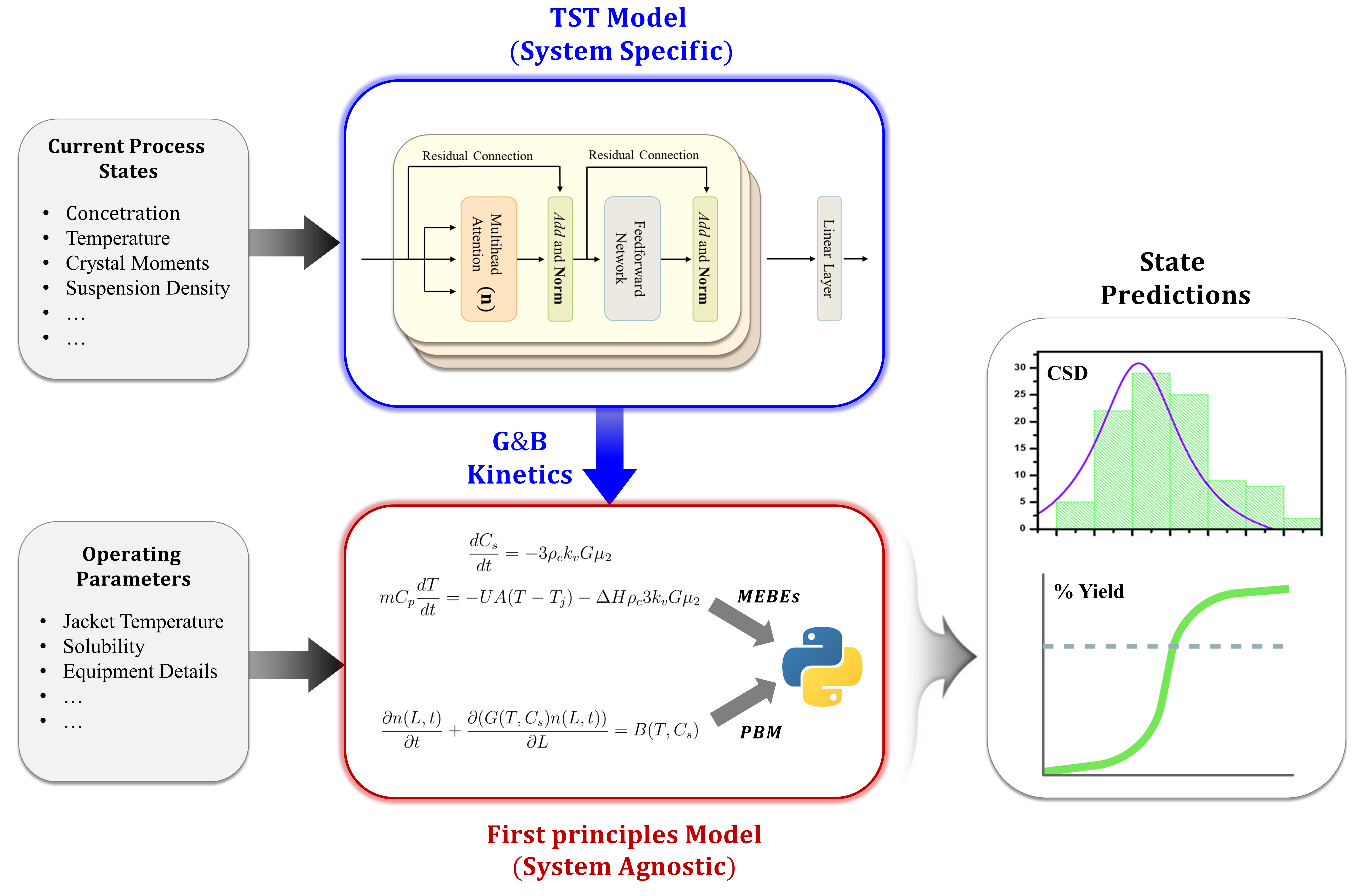}}
	\end{center}
	\caption{A schematic illustration of TST-based hybrid model in a series configuration.}
	\label{series_configuration}
\end{figure} 

\subsection{Series Hybrid Model}

Typically, a hybrid model integrates a system-agnostic FP model (e.g., PBM + MEBEs in the case of crystallization, reactor models + MEBEs in the case of fermentation, and others) with a system-specific ML model (e.g., a TST model predicting the crystallization kinetics or Arrhenius-type reaction kinetics). Moreover, in a fully-integrated, series configuration, the current system states (and other information of the operating conditions or process parameters) are fed to an ML model (in this case, a TST) to predict the system-specific kinetics for crystallization (our representative case), which is then fed to an FP model that simulates the crystallization process as illustrated in Figure~\ref{series_configuration}. Specifically, the input to the TST model is the state information for current and past $W$ time-steps (i.e., $[X_{t-W}, X_{t-W+1} ... X_{t}]$, where $ X_t = [T_j, C_s, T, \bar{L}, \mu_i, M_T, t]$) and the output of the TST model is the four kinetic parameters for $G$\&$B$ kinetics (i.e., $ \hat{z} \in [k_g, k_b, p_b, p_g]$). Next, these parameters are used in Equations \ref{G_rates} and \ref{B_rates} to compute the instantaneous value of $G$\&$B$ at time $t$, which are then fed to Equations~\ref{PBM} and \ref{MEBEs} to simulate the next-step of batch crystallization to generate state predictions ($\hat{y}$). 

These state predictions ($\hat{y}$) are then compared against the true state evolution ($y$) to compute the NMSE (i.e., $e$). During the training of the series hybrid model, the prediction errors need to backpropagate via the entire hybrid model to update the parameters ($\theta$) of the TST model using the automatic differentiation capabilities of the \textit{autograd} function embodied in the PyTorch package. Basically, PyTorch generates a computational map between all the model parameters of the TST model and the variables of the FP model, which allows it to use automatic numerical differentiation to propagate $e$ across the entire hybrid model. Briefly, $\frac{\partial e}{\partial \hat{y}}$ is computed and then backpropagated through the FP model to find the $\frac{\partial e}{\partial \hat{z}}$ (i.e, which is the output of the TST model). Once $\frac{\partial e}{\partial \hat{z}}$ is computed, \textit{autograd} performs backpropagation through each of the FFNs of the encoder and decoder blocks in a manner similar to a typical DNN to modify the parameters ($\theta$) of the TST models \cite{raissi2019physics}. More details on the training of such a hybrid model can be found in the literature \cite{bangi2020deep, shah2022deep, bangi2022physics}.

%\begin{rmk}
%	 The objective of the series hybrid model is to estimate $[k_g, k_b, p_b, p_g]$ at every time-step, which is then utilized by ODEs in the PBM and MEBEs to make state predictions that are compared with the dextrose process data. Thus, it is important that during model training that the predicted $[k_g, k_b, p_b, p_g]$ are in a certain range that does not cause major numerical instability in the ODE solver as it will result in noisy model predictions or the ODE solver may entirely fail, and in both of these cases backpropogation of errors will not be possible. That said, the advantage of a series hybrid model is that \textit{a priori} system information can be directly incorporated in the model. For instance, the supersaturation dependence on G\&B kinetics of most conventional sugar crystal systems (i.e., $p_g$ and $p_b$) is between 0 and 3 \cite{hartel1991sugar}. Similarly, by analyzing the process data, a rough estimate of the magnitude of $k_g$ and $ k_b$ can be estimated for the initial training of hybrid model. For example, batch crystallization of dextrose is performed over 20+ h, and results in crystal of average size between 100 to 200 microns, thereby suggesting $G \in [5,10]$ micron/h), which results in $k_g \in [10^{-8}, 10^{-10}]$. Similarly,  
%\end{rmk}

\begin{figure}[!ht]
	\begin{center}
		\centerline{\includegraphics[width=0.85\columnwidth]{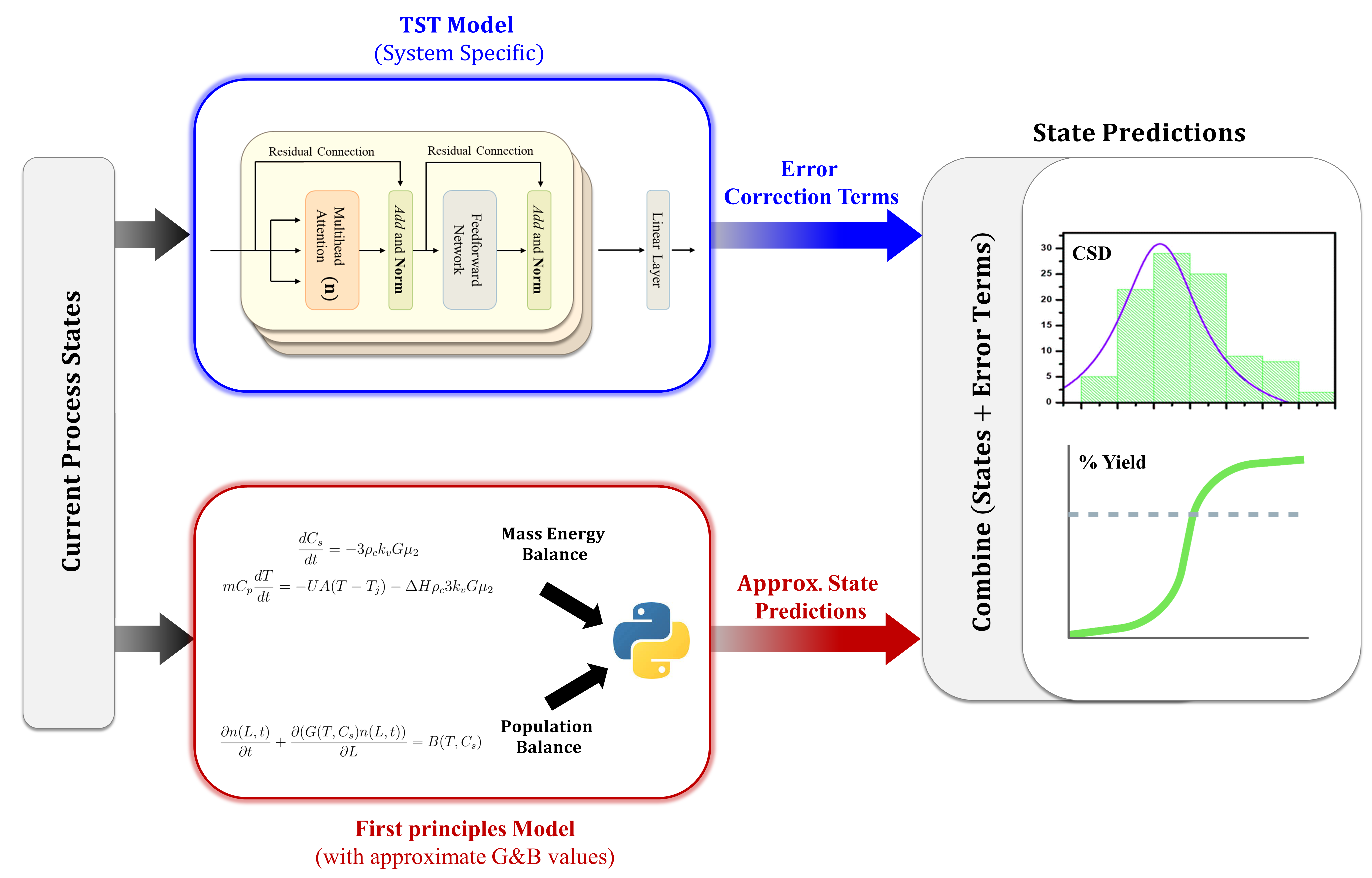}}
	\end{center}
	\caption{A schematic illustration of TST-based hybrid model in a series configuration.}
	\label{parallel_configuration}
\end{figure}

\subsection{Parallel Hybrid Model}

The parallel approach is partly similar to the series approach but with some distinct differences as shown in Figure~\ref{parallel_configuration}. The current operating conditions are used to simulate the FP model (i.e., PBM + MEBEs) using approximate $G$\&$B$ kinetics. Industrial manufacturers often introduce a new crystal product into their production chain for which the $G$\&$B$ kinetics are unknown. This presents two potential courses of action: (a) applying the known kinetics of a similar, existing system to the new one, or (b) conducting basic experimental studies to get a rough estimation of the new system's kinetics. Let's consider the first scenario. A food manufacturer that produces aspartame, a popular artificial sweetener, may decide to create a more potent variant with different crystallization kinetics. Since the kinetics of the original aspartame could be obtained from previous analyses or academic literature, this data can be leveraged to simulate the crystallization process of the new product. The predictions derived from these simulations may not perfectly align with the true system dynamics, but they do provide a general sense of the state dynamics and the temporal evolution trend of states. In the second scenario, manufacturers can conduct basic, resource-efficient experimental studies or pilot runs to gain a rough estimate of the $G$\&$B$ kinetics for the new crystal system. In either case, there will be a plant-model mismatch between the state predictions generated using the approximate kinetics and the actual outcomes. However, these approaches offer a preliminary understanding of $G$\&$B$ kinetics, which is crucial for navigating the early stages of new product development.

% For this particular case study, as explained in Section~\ref{problem_description}, batch crystallization of dextrose is considered as the desired crystal system, and it assumed that information about the approximate kinetics is known (Equation~\ref{approx_kinetics}). Since the approximate kinetics is different than true kinetics, simulating the FP model with the approximate kinetics will have a plant-model mismatch with the process data. To resolve this challenge, in a parallel hybrid model, the TST model utilizes the current process data to yield the error correction terms that can be added to the state predictions from the approximate kinetics to minimize the plant-model mismatch as shown in Figure~\ref{parallel_configuration}.

As mentioned earlier, to demonstrate the working of a parallel hybrid model, a hypothetical scenario for the case study of batch crystallization of dextrose is considered. Essentially, it is assumed that the exact kinetics of dextrose crystallization is unknown; however, an approximation of the crystallization kinetics is obtainable as described in Equation~\ref{approx_kinetics}. Utilizing the approximate kinetics in the FP model (i.e., PBM + MEBEs) will result in incorrect state predictions as the true dextrose kinetics is different (Equation~\ref{dextrose_kinetics}). These incorrect state predictions of the FP model can be combined with the error correction terms from the TST model to result in better state predictions that will minimize the plant-model mismatch. Specifically, the TST model takes in an input tensor state information for current and past $W$ time-steps (i.e., $[X_{t-W}, X_{t-W+1} ... X_{t}]$, where $ X_t = [T_j, C_s, T, \bar{L}, \mu_i, M_T, t]$) and predicts the error correction terms ($\hat{z}$) that can be added to the predicted states of the FP model ($\bar{y}$) to generate state predictions ($\hat{y}$), which will be compared against the true system dynamics values ($y$). Basically, during training of the parallel hybrid model, the NMSE between $y$ and $\hat{y}$ is the error of the parallel hybrid model that is backpropagated through the entire hybrid model using the \textit{autograd} functionality in PyTroch as described above. Briefly, $\frac{\partial e}{\partial \hat{y}}$ can be used to compute $\frac{\partial e}{\partial \hat{z}}$ (i.e, which is the output of the TST model). Next, backpropagation through each of the FFNs of the encoder and decoder blocks is used to modify the parameters ($\theta$) of the TST model. 

\begin{figure}[!ht]
	\begin{center}
		\centerline{\includegraphics[width=0.85\columnwidth]{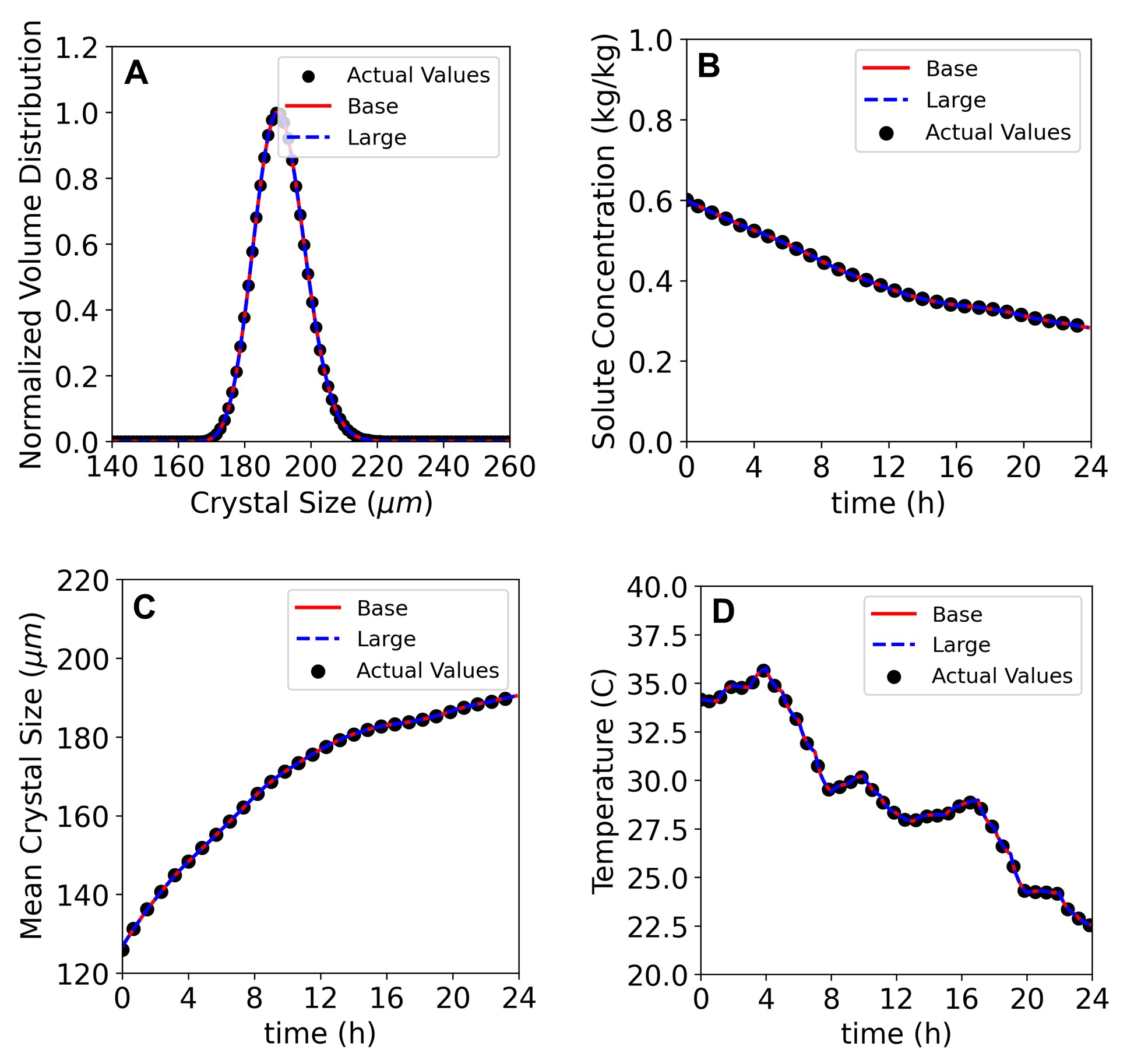}}
	\end{center}
	\caption{Comparison of predicting (a) final CSD, (b) solute concentration, (c) mean crystal size ($\bar{L}$), and (d) crystallizer temperature by the series hybrid model and true system dynamics for an arbitrarily chosen operating condition.  The $R^2$ value for all the predictions is over 0.99. }
	\label{state_evolution}
\end{figure}

\begin{rmk}
Further context can be provided for the development of two hybrid modeling approaches. First, in the parallel hybrid approach, an FP model (PBM + MEBEs) is employed, assuming an approximate version of the exact crystallization kinetics is available. However, due to discrepancies between the approximate and true kinetics, the predicted state evolution using the approximate kinetics may not align with the actual state evolution. To address this issue, a TST model is employed to predict error correction terms, which can be subsequently incorporated to rectify the predicted state evolution. Second, the approximate kinetics can be estimated by various approaches. Basically, often in the industry, a new crystal system needs to be simulated and optimized, but the exact kinetics of this new system may not be readily available. Thus, two approaches can be considered to address this limitation. The first option is to utilize the kinetics of an associated system as approximate kinetics. For instance, if the new system bears similarities to the lysozyme crystal system or represents a modified form of lysozyme, the kinetics of the traditional lysozyme system can serve as approximate kinetics for generating baseline predictions. A second but more rigorous option is to obtain a rough estimate of the kinetics via limited experimental data or by performing a few pilot studies. More precisely, an approximate value of the magnitude of growth rate, and its power law dependence on supersaturation can be estimated). Third, it is important to highlight that the parallel hybrid model differs from the series hybrid model. In the series approach, the TST model learns to predict instantaneous $G$\&$B$ kinetics directly. In contrast, the TST model in the parallel configuration aims to correct the state predictions generated using the approximate kinetics (i.e., whether derived from a similar system or through rough estimations from the literature or limited experiments). 
\end{rmk}

\section{Results and Discussion}
\subsection{Series Hybrid Model}\label{Results: Series Model}
Using the training methodology and the training dataset mentioned above, TST-based hybrid models were trained, validated, and tested as enumerated in Table~\ref{effect_of_W}. Here, the NMSE refers to plant-model mismatch, which is the deviation between predictions of the series hybrid model and the process states (i.e., NMSE = $\left\lVert \hat{y}-y \right\rVert^2 $). In addition to comparing the two architectures of the TST model (i.e., \textbf{Base} and \textbf{Large}), the effect of window size $W \in [1, 12]$ is also investigated. Basically, $W$ plays a crucial role in determining the contextual understanding and attention mechanism in TSTs. A larger window size allows the TST model to have a broader context by incorporating information from a greater number of previous time steps. This increased context enables the model to capture long-term dependencies and patterns in the time-series data, which is particularly beneficial for capturing the latent effect of process changes, delayed response to temperature changes, and other scenarios. For instance, regulating the jacket temperature in batch crystallization will have a delayed effect on the internal temperature of the crystallizer, which affects the supersaturation value that dictates $G$\&$B$ kinetics. In another scenario, rapid crystal growth in the initial period will deplete the solute concentration and will lead to low $G$\&$B$ kinetics in the later period. On the other hand, a smaller window size focuses on a narrower context and gives more weight to recent time steps. This can be advantageous when the time-series data exhibits short-term dynamics or when the most recent information is more relevant for making accurate predictions. For example, if the jacket temperature is changed more frequently (e.g., every 10 minutes), then giving more weightage to these recent time-steps will be necessary. Basically, the attention mechanism in a TST model determines the relevance and importance of different time steps within the window. It assigns higher attention weights to more informative and influential time steps while downplaying the significance of less relevant ones. The attention mechanism is crucial for capturing temporal dependencies and identifying the key patterns in the time-series data. More details on the intricacies of different window sizes can be found in the literature \cite{sitapure2023exploring,sitapure2023crystalgpt,ainslie2020etc}. 

\begin{table}[!ht]
	\centering
	\renewcommand{\arraystretch}{1.3} % Adjust the value to set row spacing
	\begin{tabular}{@{}cccc@{}}
		\toprule
		TST Model Size &            & \textbf{Base} & \textbf{Large} \\ 
		\# Parameters                                                                      &            & 2.5M          & 6.5M         \\  \midrule
		\multirow{3}{*}{\begin{tabular}[c]{@{}c@{}}NMSE ($10^{-4}$)\\ $W$ = 1\end{tabular}} & Training   & 45             & 56             \\ 
		& Validation & 81             & 73             \\ 
		& Testing    & 86             & 93             \\ 
		&            &               &              \\ 
		\multirow{3}{*}{\begin{tabular}[c]{@{}c@{}}NMSE ($10^{-4}$)\\ $W$ = 6\end{tabular}}   & Training   & 52             & 56             \\ 
		& Validation & 66             & 60             \\ 
		& Testing    & 78             & 72             \\ 
		&            &               &              \\ 
		\multirow{3}{*}{\begin{tabular}[c]{@{}c@{}}NMSE ($10^{-4}$)\\ $W$ = 12\end{tabular}} & Training   & \textbf{28}             & 25             \\ 
		& Validation & \textbf{55}             & 50             \\ 
		& Testing    & \textbf{59}             & 65              \\ \toprule
	\end{tabular}
	\caption{Performance comparison of series hybrid models. The best-performing model is highlighted in \textbf{bold}.}
	\label{effect_of_W} 
\end{table}

To this end, Table~\ref{effect_of_W} shows that NMSE values for training, validation, and testing cases are 10 to 15\% lower for the case of $W=6$, and 25 to 30\% lower for the case of $W=12$, when compared with the case for $W=1$. Here, $W=1$ refers to the case, which is typically utilized in simple DNN-based hybrid models, wherein only the current state information is utilized by the hybrid model, thereby having no inclusion of contextual information and associated trends, patterns in operating conditions, and state evolution. Also, the comparison between the \textbf{Base} and \textbf{Large} models is not significant, which suggests that a TST model with excessively large parameters is not required. Moreover, the size of the Base model is on par with the typical size of various transformer-based chemical process models available in the literature \cite{vogel2023learning, mann2021predicting, sitapure2023crystalgpt, wen2022transformers}. 

To visualize the prediction accuracy of the series hybrid model, Figure~\ref{state_evolution} shows a compilation of state predictions generated for an arbitrarily chosen operating condition. Specifically, crystal size distribution (CSD) at the end of crystallization (i.e., $t =$ 24 hr), temporal evolution of solute concentration, mean crystal size ($\bar{L}$), and crystallizer temperature predicted by the series hybrid model is compared with the true process dynamics.  It is evident from Figure~\ref{state_evolution} that the model predictions for both the \textbf{Base} and \textbf{Large} models are in excellent agreement with the true values, and showcase an $R^2$ over 0.99 for all the state variables, thereby indicating remarkable prediction accuracy. Moreover, as depicted by Table~\ref{effect_of_W}, there is very little difference between the predictive performance of the \textbf{Base} and \textbf{Large} TST model, which is reinforced by the comparison shown in Figure~\ref{state_evolution}.

\subsection{Comparison with Parallel Configuration}
As there are two configurations of hybrid models that are generally utilized in the literature, a comparison of TST-based series and parallel hybrid model is enumerated in Table~\ref{series_vs_parallel}. To have a fair comparison, the architecture of the TST model within the two hybrid models is exactly the same. It can be seen that the series hybrid model has 3 to 5 times lower NMSE values than the parallel hybrid model for both the \textbf{Base} and \textbf{Large} model cases. The variation in performance among these hybrid models can be largely attributed to the distinct roles that the TST model plays in the series and parallel configurations.

\begin{table}[!ht]
	\centering
	\renewcommand{\arraystretch}{1.2} % Adjust the value to set row spacing
	\begin{tabular}{@{}cccc@{}}
		\toprule
		TST Model Size &            & \textbf{Base} & \textbf{Large} \\ 
		\# Parameters                                                                      &            & 2.5M          & 6.5M         \\  \midrule
		\multirow{3}{*}{\begin{tabular}[c]{@{}c@{}}NMSE ($10^{-4}$)\\ \textit{Series} ($W$ = 12)\end{tabular}} & Training   & \textbf{28}             & 25             \\ 
		& Validation & \textbf{55}             & 50             \\ 
		& Testing    & \textbf{59}             & 65             \\ 
		&            &               &              \\ 
		\multirow{3}{*}{\begin{tabular}[c]{@{}c@{}}NMSE ($10^{-4}$)\\ \textit{Parallel} ($W$ = 12)\end{tabular}}   & Training   & 94             & 133             \\ 
		& Validation & 116             & 135             \\ 
		& Testing    & 372             & 403             \\ \toprule
	\end{tabular}
	\caption{Performance comparison of series and parallel hybrid model. The best-performing model is highlighted in \textbf{bold}.}
	\label{series_vs_parallel} 
\end{table}

\begin{figure}[!ht]
	\begin{center}
		\centerline{\includegraphics[width=0.85\columnwidth]{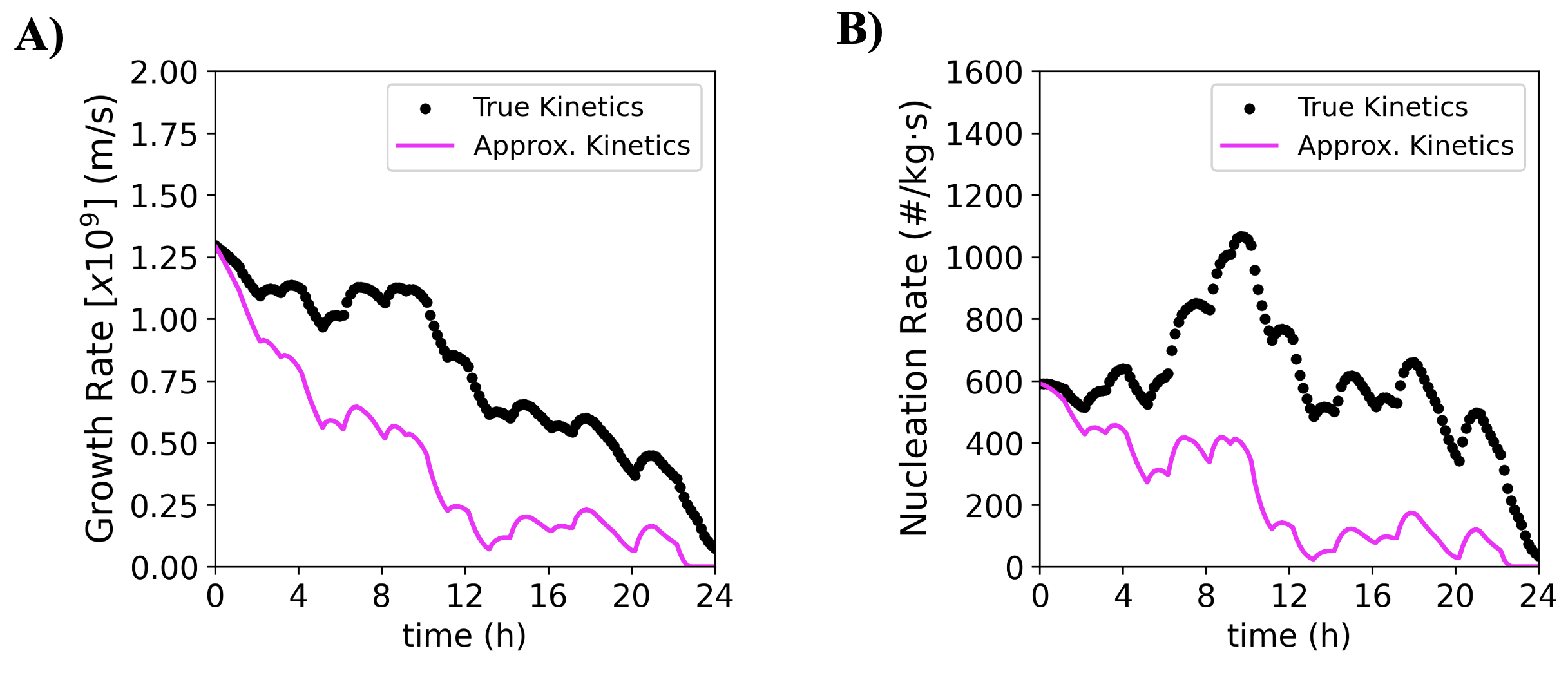}}
	\end{center}
	\caption{Comparison of true dextrose kinetics (followed by the process data), and approximate kinetics considered for the case study (Equation~\ref{approx_kinetics}).}
	\label{Approx_kinetics_copmparison}
\end{figure}

Specifically, in the parallel hybrid model, the TST model is expected to learn error correction terms that can be appended to the state predictions generated from the FP model, which utilizes the approximate kinetics (refer to Equation~\ref{approx_kinetics}). However, the difference in actual and predicted values, or the error dynamics, does not conform to a specific pattern, making it challenging for most ML models, including TSTs, to capture the inherent dynamics. For instance, with the approximate kinetics, supersaturation dependence has an exponent of 2, whereas, with the true kinetics, the growth rate has a dependence of 1 and the nucleation rate is approximately 1.5. Consequently, for a given operating condition with a certain supersaturation value, the $G$\&$B$ values will exhibit a discrepancy as shown in Figure~\ref{Approx_kinetics_copmparison}. This discrepancy between the $G$\&$B$ kinetics will subsequently introduce variations in the evolution of important process variables, such as the CSD, mean crystal size, and concentration. Therefore, the parallel configuration's TST model aims to minimize the plant-model mismatch by predicting suitable error correction terms. However, the difference between the approximate and true kinetics is dynamic and tends to fluctuate (as seen in Figure~\ref{Approx_kinetics_copmparison}). Furthermore, the FP model uses the instantaneous values of $G$\&$B$ kinetics to simulate state evolution at the subsequent time-step. Consequently, any discrepancy between the approximate and true kinetics gets propagated through a nonlinear transformation. This added complexity further heightens the challenge for the TST model in the parallel hybrid model to accurately predict the error correction terms.

% This complexity poses a challenge for most ML models, including attention-based TST models, as it becomes increasingly difficult to accurately predict the error correction terms due to the challenge of discerning the inherent dynamics within them.  

\begin{figure}[!ht]
	\begin{center}
		\centerline{\includegraphics[width=1\columnwidth]{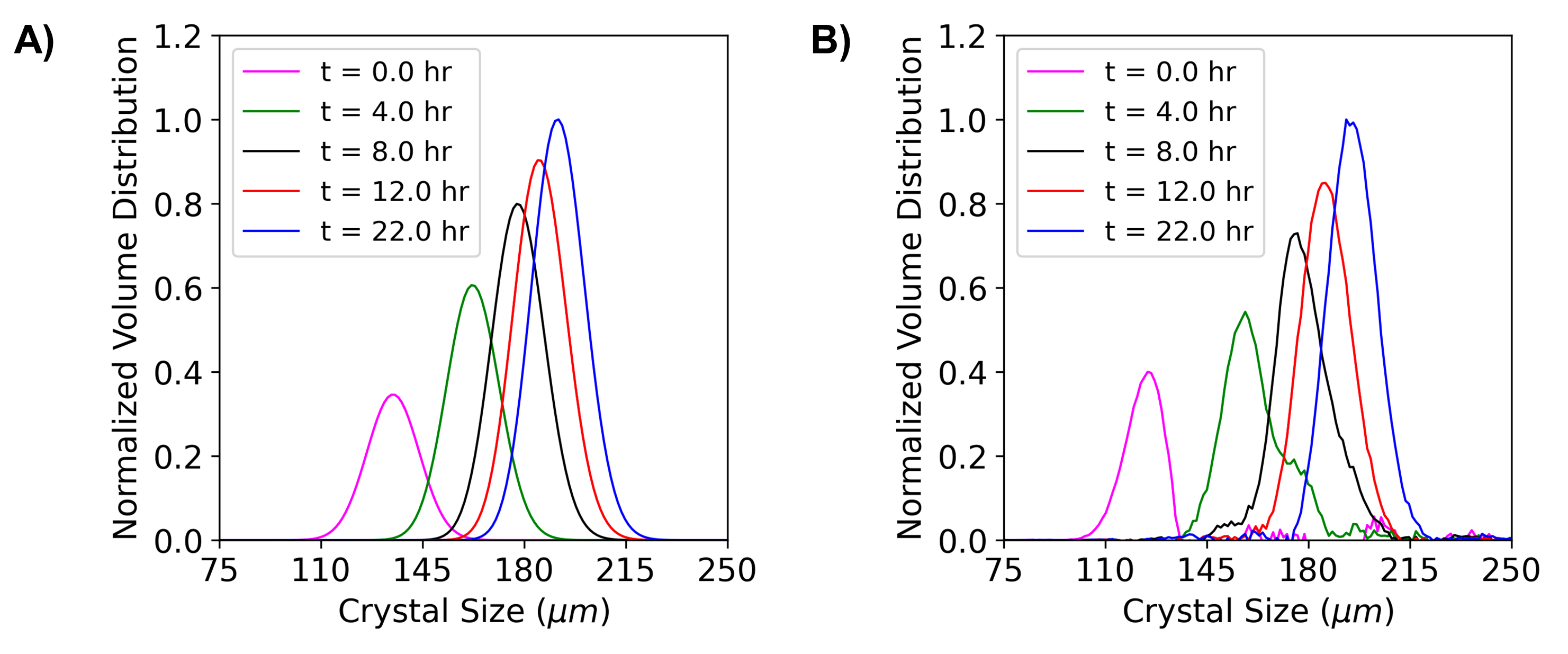}}
	\end{center}
	\caption{CSD evolution as predicted by (a) series hybrid model, and (b) parallel hybrid model. In both the cases, the TST Base model with $W=12$ is utilized. }
	\label{CSD_comparison}
\end{figure}

In contrast, the TST model in the series hybrid model is designed to learn patterns in the state dynamics and correlate them with the kinetic parameters, that is,  $\hat{z} \in [k_g, k_b, p_b, p_g]$ of the $G$\&$B$ equations. The state information from the current and previous $W$ time-steps acquired from the process data tends to follow a smoother trend compared to the error correction terms, simplifying the task of the TST model to discern the underlying interdependencies between different system states and the kinetic parameters. As the TST model progressively learns to predict the $G$\&$B$ kinetics accurately, feeding these into the FP model (PBM + MEBEs) results in state predictions that are very close to the true values. To further demonstrate these differences, Figure~\ref{CSD_comparison} shows the evolution of CSD in the batch crystallizer for an arbitrarily chosen operating condition from the testing dataset. The CSD evolution for the series hybrid model is smooth and does not demonstrate any numerical instabilities, following a typical trend where the seed crystals start with a specific distribution and mean crystal size, which continues to grow as crystallization proceeds. In contrast, Figure~\ref{CSD_comparison}b presents the CSD evolution predicted by the parallel hybrid model. Although it shows a similar trend as the series case, it exhibits numerical instabilities due to the chaotic nature of error correction terms and the associated difficulty in predicting these terms accurately. The comparison in Figure~\ref{CSD_comparison} and Table~\ref{series_vs_parallel} showcases the superior predictive ability of the series hybrid model in terms of CSD evolution. Given that the combined results from Tables~\ref{effect_of_W} and \ref{series_vs_parallel} suggest that the TST-based series hybrid model with $W=12$ shows the lowest NMSE values, this model is considered for further analysis in the manuscript. 

\subsection{Model Interpretability}
Apart from the performance difference between the series and parallel hybrid models, there are differences in the utility and interpretability of these models. Specifically, the parallel hybrid model predicts the error correction terms that need to be added to the FP model (inclusive of the approximate kinetics) to minimize plant-model mismatch. The TST model within the parallel hybrid model, thereby acts as a black box model between the current process states and the error correction terms, thereby leading to very poor interpretability. On the contrary, the fully-integrated, series hybrid model directly predicts instantaneous $G$\&$B$ kinetics for every time-step, which are then utilized by the FP model to predict the state evolution. In this scenario, along with procuring accurate state predictions, the instantaneous time-varying $G$\&$B$ values can be predicted as well, thereby providing a computational probe into the kinetics of a desired crystal system. 

\begin{figure}[!ht]
	\begin{center}
		\centerline{\includegraphics[width=0.75\columnwidth]{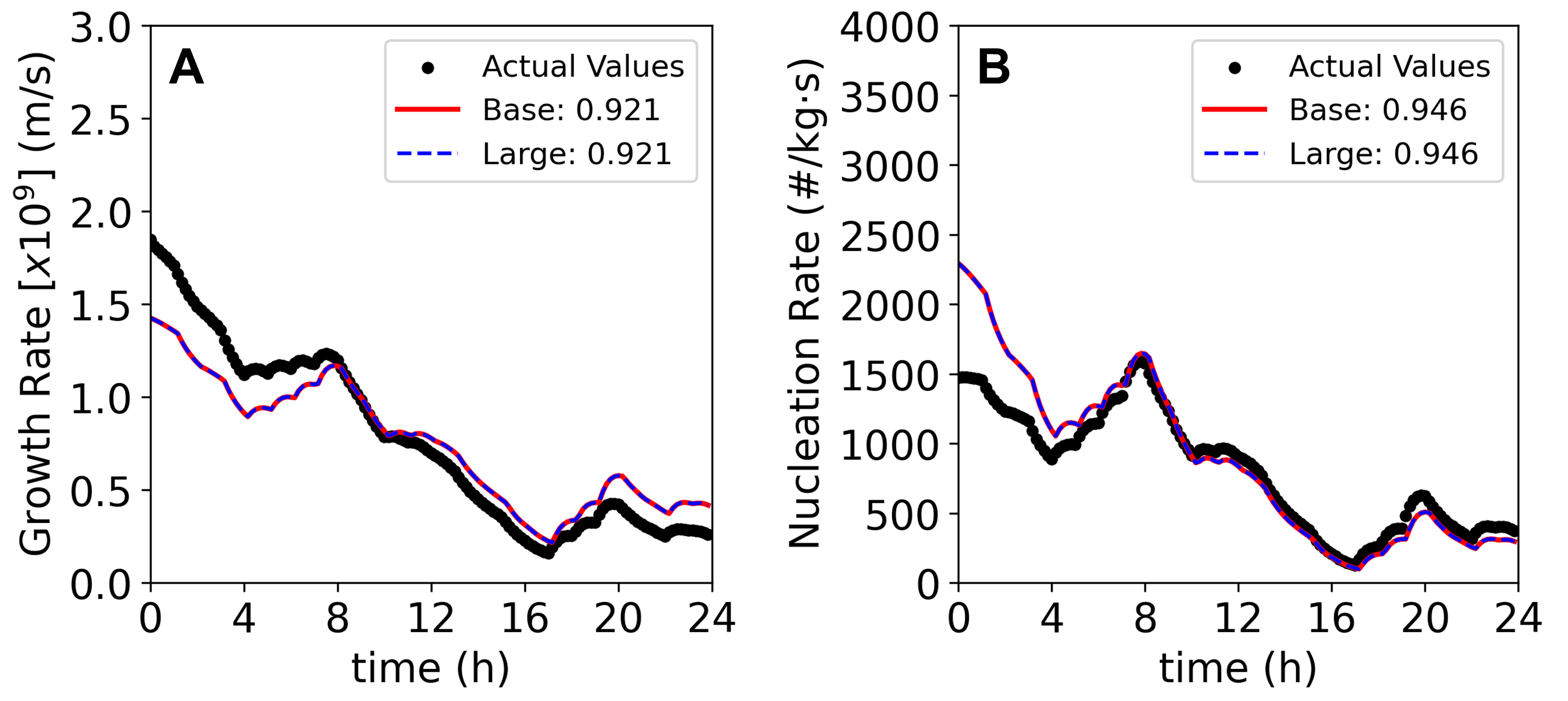}}
	\end{center}
	\caption{Comparison of predicted $G$\&$B$ kinetics and true kinetics for the series hybrid model. }
	\label{G&B_predictions}
\end{figure}

For instance, Figure~\ref{G&B_predictions} showcases the instantaneous growth and nucleation rates for an arbitrary operating condition predicted by the series hybrid model along with their comparison to true dextrose crystallization kinetics (Equation~\ref{dextrose_kinetics}). It is evident that the $G$\&$B$ predictions are in very good agreement with practical values, and showcase $R^2$ values of 0.90+. It is interesting to note that the major deviation between  $G$\&$B$ predictions is during the start of crystallization (i.e., $t =[0,4]$ hr), and after that, the $G$\&$B$ values follow the real values very closely. This observation is attributed to the working of attention-based TST models. More specifically, when more process data is available for the TST model, it can intelligently leverage its attention mechanism, which is adept in understanding the contextual interdependencies between the system states and outputs (i.e., $G$\&$B$ kinetics in this case) as explained in Section~\ref{Results: Series Model}.  Thus, as the crystallization progresses, and we collect more process data (i.e., $T_j$, $C_s$, $T$, $\mu_i$, etc.), the TST is proficient in recognizing patterns and trends between not just singular states, but also their interdependencies through the attention mechanism \cite{sitapure2023crystalgpt}, thereby leading to more accurate $G$\&$B$ predictions after $t=4$ hr.

\begin{figure}[!ht]
	\begin{center}
		\centerline{\includegraphics[width=0.75\columnwidth]{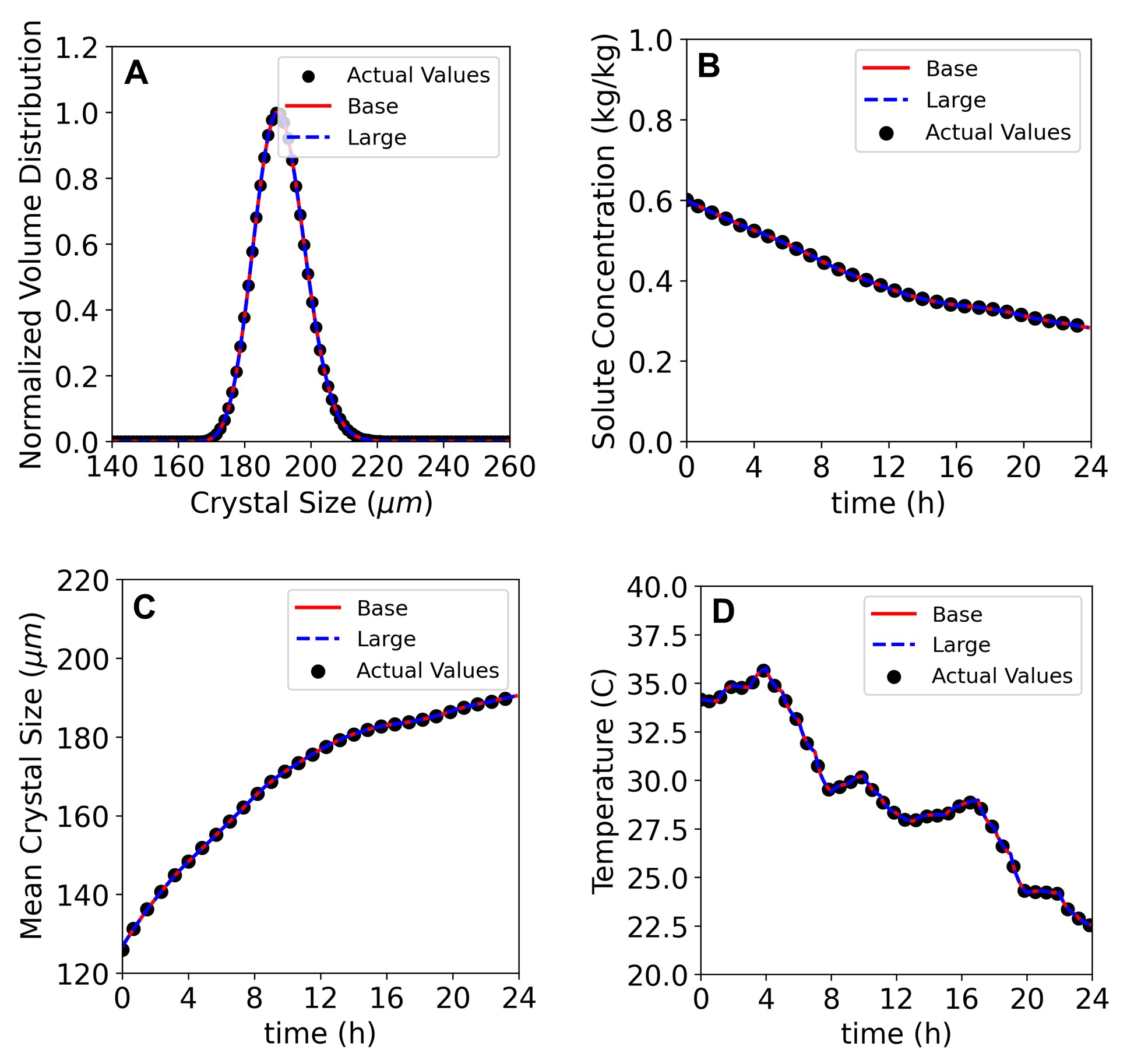}}
	\end{center}
	\caption{Comparison of predicting (a) final CSD, (b) solute concentration, (c) mean crystal size ($\bar{L}$), and (d) crystallizer temperature by the series hybrid model and true system dynamics under the same operating condition as in Figure~\ref{G&B_predictions}.  The $R^2$ value for all the state predictions is over 0.99. }
	\label{Case 2: State_evolution}
\end{figure}

That said, it is remarkable that the series TST-hybrid model can accurately predict not just a single value for $G$\&$B$ kinetics, but trace temporal evolution along the crystallization process. More importantly, although $G$\&$B$ show predictions with an $R^2 \in [0.90,0.95]$ instead of a perfect match (i.e., an $R^2$ value over 0.99), their effect on the evolution of the system states in batch crystallization is not impacted by this small discrepancy. This can be explained as follows: First, the FP model comprises a complex set of ordinary differential equations (ODEs) for modeling the population dynamics, energy balance at the jacket wall, the depletion of solute concentration due to nucleation and growth, the effect of temperature-dependent solubility, and other factors. Second, these various ODEs utilize $G$\&$B$ values in conjunction with a plethora of other variables that dampen the effect of slightly different $G$\&$B$ values for that time-step. In simpler terms, a $\delta$ deviation in predicting $G$\&$B$ kinetics does not drastically affect the accuracy of the state predictions (i.e., CSD evolution, concentration profile, temperature profile, and others). To further demonstrate this aspect, Figure~\ref{Case 2: State_evolution} shows the prediction of state evolution for the same operating condition that was utilized for generating Figure~\ref{G&B_predictions}. It is evident that the corresponding state predictions are in excellent agreement with the true system dynamics, and showcase $R^2$ values over 0.99 for all state predictions. 

\subsection{Discussion}
The combination of the aforementioned simulation results and comparison between series and parallel TST-based hybrid models provide three key insights. First, in terms of prediction accuracy, the series configuration is $\sim$ 5 times more accurate than the parallel configuration. As mentioned earlier, this is attributed to the remarkable ability of the attention-based TST mechanism to find the underlying interdependencies between different system states and the output (i.e., $G$\&$B$ kinetic parameters) in the case of the series hybrid model. However, the tumultuous nature of the error correction terms, which makes it difficult to recognize distinctive patterns, results in the poor performance of the parallel hybrid model. Second, the series hybrid model has high interpretability as (a) it allows accurate estimation of instantaneous $G$\&$B$ kinetics, and (b) allows direct integration of \textit{a priori} kinetics (e.g., the structure of the growth and nucleation rate equations, and their dependence on concentration and temperature). Since the parallel hybrid model learns to predict the error correction terms to amend the state predictions from the FP model (inclusive of the approximate kinetics), it acts as a black box model, thereby having very poor interpretability. Third, the interpretability of the series hybrid model also empowers it to be used as a modular package that can be integrated into various different crystallization configurations. For instance, in the current case study, the TST predicts the $G$\&$B$ kinetics, which are fed to the FP model of a batch crystallizer. However, the same TST model can be integrated with an FP model for a continuous tubular crystallizer or a continuously agitated crystallizer \cite{kwon2014crystal, kwon2014enhancing, SITAPURE2020127905, sitapure2021cfd}. Basically, given the internal configuration of the series hybrid model, it is possible to replace the FP model with another FP crystallization model, and the new series hybrid model can be easily fine-tuned for a certain case. On the other hand, since the TST in the parallel series model predicts the error correction term that needs to be added to the FP model, if the FP model changes, the entire hybrid model will require retraining. Overall, the series hybrid model demonstrates superior performance, greater interpretability, and higher modularity as compared to the parallel hybrid model configuration for developing hybrid model-based digital twins. 

\section{Conclusions}
In the era of next-generation digital twins, the development of hybrid models that combine physics-based principles with advanced machine learning has sparked immense interest. In this exciting work, cutting-edge attention-based TSTs are seamlessly integrated with a first-principles (FP) model for complex batch crystallization systems. Leveraging the power of multi-headed attention and positional encoding, TSTs exhibit remarkable predictive performance by capturing both short-term fluctuations and long-term trends in process dynamics. To thoroughly evaluate the potential of TST-based hybrid models, two distinct configurations, series, and parallel, are meticulously compared across a wide range of operating conditions. The series approach emerges as the clear winner, boasting a striking five-fold reduction in NMSE. This impressive performance can be attributed to the TST's extraordinary ability to recognize intricate interdependencies among system states, enabling accurate prediction of $G$\&$B$ kinetics. Moreover, the series hybrid model consistently achieves exceptional $R^2$ values exceeding 0.99 across diverse operating conditions. The series configuration also excels in interpretability, enabling accurate estimation of temporal predictions for the critical $G$\&$B$ kinetics. In contrast, the parallel hybrid model relies on error correction terms and fails to provide direct insight into $G$\&$B$ values. These findings underscore the series hybrid model's superior performance, greater interpretability, and higher modularity, making it the ideal choice for developing hybrid model-based digital twins. With the rising adoption of digital twins, attention-based hybrid models like the series TST-based approach are poised to revolutionize the landscape of chemical manufacturing. 

\section{Acknowledgments}
Financial support from the Artie McFerrin Department of Chemical Engineering, and the Texas A\&M Energy Institute is gratefully acknowledged.

\newpage

%Bibliography
\bibliographystyle{unsrt}  
\bibliography{HybridTST_bib}  
\end{document}